\documentclass[prl,aps,twocolumn]{revtex4}
\usepackage{amsthm}
\usepackage{amsfonts}

\begin{document}

\title{Can Quantum Cryptography Imply Quantum Mechanics? \\
  Reply to Smolin}

\author{Hans Halvorson}

\address{Department of Philosophy, Princeton University, \\ 
Princeton, NJ 08544 hhalvors@princeton.edu }

\author{Jeffrey Bub}

\address{Department of Philosophy, University of Maryland, \\ 
College Park, MD 20742 jbub@carnap.umd.edu} 

\date{\today}

\begin{abstract}
  Clifton, Bub, and Halvorson (CBH) have argued that quantum mechanics
  can be derived from three cryptographic, or broadly
  information-theoretic, axioms.  But Smolin disagrees, and he has
  given a toy theory that he claims is a counterexample.  Here we show
  that Smolin's toy theory violates an independence condition for
  spacelike separated systems that was assumed in the CBH argument.
  We then argue that any acceptable physical theory should satisfy
  this independence condition.
\end{abstract}

\maketitle

\section{Introduction}

In a recent note, Smolin \cite{smolin} has presented a toy theory that
simulates some interesting cryptographic features of quantum
mechanics.  Most interestingly, Smolin's toy theory satisfies the
three cryptographic, or information-theoretic, axioms from which
Clifton, Bub, and Halvorson (CBH) \cite{cbh} have claimed to be able
to derive quantum mechanics.  So, Smolin argues that, \emph{contra}
CBH, QM cannot be derived from these three axioms.

We agree with Smolin that QM is not a \emph{logical} consequence of
the three information-theoretic axioms, taken in complete isolation
from any theoretical context.  In fact, we think that attempting such
a derivation would be futile, as shown by the history of failed
attempts (e.g., the quantum logic program) to derive QM from
completely explicit, physically plausible axioms.  When such attempts
have not failed miserably, their partial successes have come at the
expense of complicating the axioms to the point of destroying all
physical insight.

The failure of attempts at theoretically-neutral derivations of QM
does not undermine the importance of providing characterizations
within some judiciously chosen framework of background assumptions ---
these assumptions might be explicit (as in CBH's assumption that
theories permit a $C^{*}$-algebraic formulation), or they might be
tacit (as, e.g., in Einstein's assumption that spacetime is continuous
and not discrete).  For someone concerned with diachronic
relationships between theories, it is an extremely interesting
question to ask whether there is a framework that encompasses both the
old and the new theory, and whether there are salient physical
postulates that distinguish the two theories.  CBH have answered this
question in the affirmative for classical and quantum mechanics: the
$C^{*}$-algebraic framework encompasses both theories, and quantum
mechanics is distinguished in terms of its satisfaction of the three
information-theoretic axioms.

But to be more specific, we argue here that Smolin's toy theory is so
remote from classical or quantum mechanics that it holds little
physical interest.  In particular, we show that Smolin's theory
violates an independence condition for distinct systems that is taken
for granted in both classical and quantum mechanics.  We then argue
that the failure of this independence condition leads to pathologies
that are unacceptable in any physical theory.

\section{Against Serial Numbers}

CBH argue that QM can be derived from three axioms: no superluminal
information transfer via measurement, no cloning \footnote{More
  precisely, no broadcasting, which reduces to no cloning for pure
  states.}, and no bit commitment.  Roughly speaking, the no cloning
axiom says that there cannot be a machine that accepts arbitrary input
states, and returns two copies of any state it receives.  The no bit
commitment axiom states that it is not possible for one observer,
Alice, to send a bit value to a second observer, Bob, in such a way
that Bob cannot access the bit value until Alice provides him with a
key, and such that Alice cannot change her bit value after she has
sent it to Bob.  It is well-known that elementary QM satisfies these
three cryptographic axioms.  CBH claim that QM can also be derived
from these three axioms, and so the conjunction of the axioms is
equivalent to the claim that QM is true.

Smolin's toy theory consists of symmetric pairs of lockboxes, where
each pair of lockboxes has a unique serial number.  Furthermore, each
lockbox contains a bit value, which is accessible to inspection only
when the lockbox is in the presence of its partner.  For the details
of how the lockbox theory satisfies the three axioms, we refer the
reader to Smolin's paper.  But note that the assumption of unique
serial numbers is needed to ensure that cloning is impossible.

Most of the details of Smolin's lockbox theory are irrelevant to his
argument against the CBH characterization result.  Indeed, in his
final discussion Smolin claims that there is a \emph{trivial}
counterexample to the CBH result.
\begin{quote} \dots there is a trivial theory that satisfies the other
  axioms [i.e., the three axioms of the CBH characterization argument].
  Namely, a theory with only one type of element, a box with a unique
  serial number and no bit value inside at all. Such a box cannot be
  cloned or broadcast, due to the serial number, cannot communicate
  superluminally, and cannot be used for bit commitment.  (p.~4)
\end{quote}
To ``make the question interesting,'' Smolin proposes a fourth
cryptographic axiom (viz., the possibility of unconditionally secure
key distribution), and his lockbox theory is intended to show that QM
does not follow from the four axioms.  However, we maintain that
neither the trivial theory nor the more sophisticated lockbox theory
stand as counterexamples to the claim that QM can be derived ---
within a completely reasonable framework --- from the three
cryptographic or information-theoretic axioms.  So, we will henceforth
ignore the key distribution axiom.

Smolin's trivial, one lockbox theory is a special case of the
following construction: consider any classical theory $T$, with any
finite number of physical objects, and where all objects travel
subluminally.  Since $T$ has no ambiguous mixtures, $T$ does not allow
bit commitment.  However, in the absence of further constraints, $T$
permits cloning of arbitrary pure states \cite[Theorem~2]{cbh}.  So,
we modify $T$ by stipulating that cloning is impossible; we can do
this by adding a law which states that there is at least one object
that has a property which no other object can have.  (In Smolin's
theory, $P$ is the serial number of a lockbox pair.)  Let's call the
resulting theory $T'$.

The theory $T'$ satisfies the three axioms of the CBH argument, but
$T'$ is clearly not quantum mechanics.  It would be natural, then, to
conclude that the failure of the CBH theorem to rule out $T'$ is due
to the further assumption that physical theories permit a
$C^{*}$-algebraic formulation.  Although this may be true, we shall
show that $T'$ violates a more fundamental independence condition for
distinct systems, and so should not be taken seriously as a physical
theory.

\section{The Schlieder Condition}

Let $A$ and $B$ denote, respectively, Alice and Bob's systems, and let
$AB$ denote the composite.  At present, we make no assumptions
concerning the mathematical structure of Alice and Bob's state spaces,
or about the means for constructing the state space of $AB$ from the
state spaces of $A$ and $B$.  Consider now the following three
independence conditions:
\begin{enumerate}
\item (Schlieder Condition) For any states $x$ of $A$, and $y$ of $B$,
  there is a state $z$ of $AB$ with marginals $x$ and $y$.  (This is a
  well-known condition from axiomatic field theory; see, e.g.,
  \cite{summers}.)
\item (Duplication of a Known State) For any state $x$ of $A$, there is a
  machine $M_{x}$ that prepares a state $z$ of $AB$ with marginals $x$
  and $x$.  (Note the order of the quantifiers.)
\item (Independent Preparation) Alice's ability to prepare states is
  independent of the state of Bob's system.
\end{enumerate}
It is clear that all three of these conditions are satisfied by the
standard Cartesian product representation of composite systems in
classical mechanics, as well as by the standard tensor product
representation of composite systems in quantum mechanics.  CBH also
assume a version of the Schlieder condition (viz.,
$C^{*}$-independence) in their derivation of QM \cite[p.~1574]{cbh}.
However, Smolin's theories violate each of these conditions.  In
particular, although there is a state of $A$ in which Alice has a pair
of lockboxes with serial number $s$, and there is a state of $B$ in
which Bob has a pair of lockboxes with serial number $s$, these states
are mutually incompatible.

The failure of the Schlieder condition entails that Alice can acquire
information about Bob's system just by determining her own state (even
if she has no prior information about their joint state).  In
particular, if Alice determines that her state is $x$, and if $y$ is
one of Bob's states that is incompatible with $x$, then Alice knows
that Bob's system cannot possibly be in state $y$.  But if $A$ and $B$
are distinct (e.g., if they are spacelike separated), then information
about the state of $A$ should not, by itself, provide information
about the state of $B$.  So, there is reason to think that in Smolin's
theory, we are not really dealing with distinct physical systems.

The failure of the Independent Preparation condition entails that
there will be mysterious constraints on which states Alice and Bob can
prepare.  For example, if there is no state with marginals $x$ and
$y$, and if Alice's system is in state $x$, then Bob will be
frustrated in his attempts to prepare $y$.  Once again, however, if
$A$ and $B$ are independent systems (e.g., if they are spacelike
separated), what physical mechanism could explain Bob's inability to
prepare $y$?

The failure of the Duplication Condition entails that there are states
that no experimenter can duplicate, even if he is supplied
\emph{every} bit of information about that state, and even if he has
unlimited physical resources.  For example, in Smolin's theory, it is
impossible to duplicate a lockbox pair --- although we are given no
explanation of what would prevent the experimenter from achieving this
goal.  This prohibition on the duplication of states is much stronger
than the ordinary quantum mechanical prohibition on cloning unknown
states.  In QM, while there is no machine that duplicates arbitrary
(unknown) input states, there is, for each fixed state $x$, a machine
$M_{x}$ that duplicates $x$.  (Let $y$ be the ready state of the
machine, and let the operation of the machine be given by the mapping
$I\otimes U$, where $U$ is a unitary operator mapping $x$ to $y$.)
Thus, cloning is impossible in QM for a very different reason from why
cloning is impossible in Smolin's theory --- in QM, the no cloning
theorem is a non-trivial result, whereas in Smolin's theory it results
from an {\it ad hoc} stipulation that states cannot be duplicated.

Finally, the failure of the Schlieder condition also raises problems
for giving an account of measurement interactions.  In particular,
suppose that $Q$ is an observable that can take finitely many values
$q_{1},\dots ,q_{n}$, and suppose that $x_{1},\dots ,x_{n}$ are states
in which $Q$ definitely has the corresponding value.  An ideal
measurement of $Q$ is normally defined as an interaction that
perfectly correlates the states $x_{1},\dots ,x_{n}$ of the object
with states $y_{1},\dots ,y_{n}$ of some measuring apparatus.
However, if the Schlieder condition fails, then there is no guarantee
that the posterior states exist, and it becomes unclear how to
formulate a general notion of measurement.

\section{Haecceities and Teleportation}

A theory is \emph{haecceitist} if it stipulates that each object has a
certain unique `thisness' or \emph{haecceity} that distinguishes it
from all other objects --- over and above the totality of its
properties that specifies its `whatness' or \emph{quiddity}.  (We
borrow this terminology from metaphysics; see, e.g., \cite{stanford}.)
So, for example, Smolin's toy theory is haecceitist because lockbox
pairs have unique serial numbers.  We are unaware of any successful
physical theory in the past, say, 400 years that has been haecceitist
in this sense.  But rather than appeal to history, we can rule out
haecceitistic theories by noting that, in a heacceitistic theory,
either teleportation is excluded by fiat because states with unique
haecceities or identifying properties cannot be prepared (they are
simply declared to be `read only'), or superluminal signaling is
possible. Since we assume that superluminal signaling is impossible,
we rule out haecceitistic theories because we wish to consider
theories in which teleportation is at least prima facie possible ---
i.e., we want to come to an understanding of the physical grounds for
the possibility or impossibility of teleportation.

To see that haecceitist theories cannot satisfy both teleportation and
no superluminal signaling, consider first a theory in which particles
can be created.  Then Alice and Bob can agree to the following
protocol for sending a bit of information superluminally: Alice and
Bob start out together and note the unique identifying properties $P$
and $Q$ of two particles.  Alice goes off with these particles while
Bob tries, alternating once every second, to create a particle with
property $P$ (respectively, property $Q$).  In order to signal value
$0$ to Bob, Alice destroys her $P$ particle, and Bob becomes aware of
this fact within one second, because he can create a particle with
property $P$.  In order to signal value $1$ to Bob, Alice destroys her
$Q$ particle, and again Bob becomes aware of this fact within one
second.

Suppose now that particles cannot be created.  Then if Bob is holding
a particle with unique identifying property $P$, the only way for
Alice to obtain a particle with property $P$ is for Bob to send his
particle to her.  So, either the particle must travel superluminally,
or the particle's state cannot be teleported.  Therefore, regardless
of whether or not particles can be created, haecceitist theories
cannot have teleportation without superluminal signaling.

\section{Conclusion}

We have seen that Smolin's theory violates the Schlieder condition,
and so has a number of physical pathologies: Alice can gain
information about Bob's system by making measurements on her system;
there will be inexplicable constraints on Alice and Bob's ability to
prepare states; and it is impossible to duplicate known states.  But
even if we ignore these pathologies, Smolin's theory could be ruled
out by requiring that the theories under consideration should not
include assumptions that, in the context of the information-theoretic
axioms, preclude the possibility of teleportation.  We conclude from
all of these facts that Smolin's theory is so different from the
theories we know (viz., classical and quantum mechanics) that it need
not be taken as a serious physical possibility.
  
Recently, Spekkens \cite{spekkens} has constructed a toy theory ---
for a different purpose --- which could also be used to challenge the
CBH argument.  In particular, Spekkens' toy theory satisfies the three
cryptographic axioms of the CBH argument, but it admits a local hidden
variable model, and so is inconsistent with QM. But, unlike Smolin's
theory, Spekkens' theory does satisfy the Schlieder condition (see pp.
17, 20).  We can conclude (by applying the contrapositive of the CBH
theorem) that Spekkens' theory does not admit a $C^{*}$-algebraic
formulation.  In fact, a stronger claim is proven in \cite{halvo}:
neither the state space of Spekkens' theory (which does not allow
arbitrary mixtures), nor the convex extension of the Spekkens state
space, is the state space of a Jordan-Banach algebra. The JB-algebraic
framework is more general than the $C^{*}$-algebraic framework and
includes a very broad class of theories in which the state space has a
convex structure. But we do not think that representability within the
$C^{*}$-algebraic framework or JB-algebraic framework is a necessary
condition for physical possibility; and so the status of Spekkens' toy
theory {\it vis a vis} the CBH characterization argument should be
examined in more depth to settle this issue.

\end{document}